\documentclass[prl,twocolumn,twoside,showpacs,floatfix,superscriptaddress]{revtex4}

\usepackage{here}

\usepackage{graphicx,color,rotating}
\usepackage{amsmath,amssymb}
\usepackage{dcolumn}
\usepackage{times,txfonts}
\usepackage{pifont}
\usepackage{multirow}

\usepackage{mathrsfs}

\usepackage{dsfont}

\usepackage{pstricks}
\usepackage{pst-text}
\usepackage{pst-coil}
\usepackage{pst-node}
\usepackage{pstricks-add}

\definecolor{FGViolet}{rgb}{0.61,0.32,0.61}
\definecolor{FGDarkBlue}{rgb}{0,0,0.6}
\definecolor{FGBlue}{rgb}{0,0,0.8}
\definecolor{FGLightBlue}{rgb}{0.2, 0.6, 0.8}
\definecolor{FGGreen}{rgb}{0.2,0.7,0.2}
\definecolor{FGLightGreen}{rgb}{0.4,1,0.4}
\definecolor{FGYellow}{rgb}{1,0.95,0}
\definecolor{FGOrange}{rgb}{0.95,0.5,0.1}
\definecolor{FGRed}{rgb}{0.8,0,0}
\definecolor{FGWhite}{rgb}{1,1,1}
\definecolor{FGLightGray}{rgb}{0.8,0.8,0.8}
\definecolor{FGGray}{rgb}{0.5,0.5,0.5}
\definecolor{FGDarkGray}{rgb}{0.3,0.3,0.3}
\definecolor{FGBlack}{rgb}{0,0,0}

\newcommand{\elem}[2]{\ensuremath{{}^{#2}\text{#1}}}
\newcommand{\emax}{\ensuremath{e_{\text{max}}}}
\newcommand{\EEEmax}{\ensuremath{E_{\rm 3max}}}

\newcommand\ESRG{E_{\rm SRG}}
\newcommand\rampFourtyC{\ensuremath{\mathcal{A}}}
\newcommand\rampFourtyJ{\ensuremath{\mathcal{B}}}
\newcommand\rampFourtyK{\ensuremath{\mathcal{C}}}
\newcommand\rampFourtyL{\ensuremath{\mathcal{D}}}

\newcommand\hw{\hbar \Omega}
\newcommand\hwSRG{\hbar \Omega_{\rm SRG}}
\newcommand\NNN{\emph{NN}+3\emph{N}}

\begin{document}


\title{\emph{Ab Initio} Path to Heavy Nuclei}

\author{Sven Binder}
\email{sven.binder@physik.tu-darmstadt.de}
\affiliation{Institut f\"ur Kernphysik, Technische Universit\"at Darmstadt, 64289 Darmstadt, Germany}

\author{Joachim Langhammer}
\affiliation{Institut f\"ur Kernphysik, Technische Universit\"at Darmstadt, 64289 Darmstadt, Germany}

\author{Angelo Calci}
\affiliation{Institut f\"ur Kernphysik, Technische Universit\"at Darmstadt, 64289 Darmstadt, Germany}

\author{Robert Roth}
\affiliation{Institut f\"ur Kernphysik, Technische Universit\"at Darmstadt, 64289 Darmstadt, Germany}

\date{\today}

%
\begin{abstract}  
We present the first \emph{ab initio} calculations of nuclear ground states up into the domain of heavy nuclei, spanning the range from \elem{O}{16} to \elem{Sn}{132}, based on two- plus three-nucleon interactions derived from chiral effective field theory. 
We employ the similarity renormalization group for preparing the Hamiltonian and use coupled-cluster theory to solve the many-body problem for nuclei with closed sub-shells.
Through an analysis of theoretical uncertainties resulting from various truncations in this framework, we identify and eliminate the technical hurdles that previously inhibited the step beyond medium-mass nuclei,
allowing for reliable validations of nuclear Hamiltonians in the heavy regime.
Following this path we show that chiral Hamiltonians qualitatively reproduce the systematics of nuclear ground-state energies up to the neutron-rich Sn isotopes.

\end{abstract}

\pacs{21.30.-x, 05.10.Cc, 21.45.Ff, 21.60.De}

\maketitle


%
\paragraph{Introduction.}
Hamiltonians derived within chiral effective field theory~\cite{MaEn11,EpHa09} represent a milestone in the endeavor to describe nuclear properties in a universal framework based on QCD.
Already at the current stage, chiral two-nucleon (\emph{NN}) plus three-nucleon (3\emph{N}) Hamiltonians have  successfully been applied in a wide range of \emph{ab initio} nuclear structure~\cite{HaPa10,HaHj12,BaBa13,HeBi13} and reaction calculations~\cite{HuLa13}.
Particularly the medium-mass regime has seen amazing progress over the past few years,---several \emph{ab initio} many-body methods can nowadays access this regime.
The importance truncated no-core shell model~\cite{Roth09,RoCa14} provides quasi-exact solutions that serve as benchmark points for computationally efficient medium-mass methods~\cite{HeBi13}.
In addition to its success in quantum chemistry, coupled-cluster theory~\cite{HaPa10,HaHj12} has emerged as one of the most efficient and versatile tools for the accurate computation of (near-)closed-shell nuclei.
Alternative approaches are the self-consistent Green's function methods~\cite{SoDu11,CiBa13,SoCi14} and the in-medium renormalization group~\cite{HeBo13,HeBi13}, which also have been generalized to open-shell systems.
Extending the range of such calculations to heavier nuclei provides important information about 
whether \emph{ab initio} methods based on universal chiral interactions are capable of describing heavy nuclei.
While most of the many-body methods above can be applied to heavier systems, challenges regarding the preparation of the Hamiltonian have prevented \emph{ab initio} theory from entering this mass range so far.

In this Letter we overcome these limitations and present \emph{ab initio} calculations of nuclei up to \elem{Sn}{132} using similarity renormalization group (SRG)-transformed chiral \NNN{} interactions. We present key developments in the treatment of the Hamiltonian that enable these calculations, and discuss the remaining uncertainties due to truncations.
For the solution of the many-body problem we use coupled-cluster (CC) theory including a non-iterative treatment of triply excited clusters.
%


\paragraph{Preparation of the Hamiltonian.}
With \emph{ab initio} nuclear structure theory advancing towards heavier systems, the preparation of the \NNN{} Hamiltonian prior to the many-body calculations becomes increasingly important. 
We start from the chiral \emph{NN} interaction at $\rm N^3LO$~\cite{EnMa03} and a local form of the chiral 3\emph{N} interaction at $\rm N^2LO$~\cite{Navr07} with regulator cutoff of \mbox{$\rm 400 \, MeV/c$}~\cite{RoLa11,RoBi12,RoCa14}.
To enhance the convergence behavior of the many-body calculations, we soften this initial Hamiltonian through a SRG transformation, formulated as flow equation in terms of a continuous flow parameter $\alpha$~\cite{BoFu07,JuNa09,RoNe10,RoLa11}.
The SRG allows to consistently evolve the \emph{NN} and 3\emph{N} interactions~\cite{RoCa14} and yields a model-space independent Hamiltonian.
One of the challenges are the many-body interactions induced during the SRG flow.
For practical reasons we truncate these interactions at the 3\emph{N} level and consequently violate the unitarity of the transformation, which introduces a flow-parameter dependence of observables. 
This $\alpha$-dependence carries information about the relevance of omitted many-nucleon interactions and allows  conclusions about their origins and importance.
We consider two types of Hamiltonians in order to distinguish the effects of the initial chiral 3\emph{N} interaction from SRG-induced contributions:
for the \NNN-induced Hamiltonian we start from the chiral \emph{NN} interaction and keep induced interactions up to the 3\emph{N} level, whereas for the \NNN-full Hamiltonian we start with the chiral \NNN{} interaction and keep all 3\emph{N} contributions.
Due to their enormous number, an energy truncation $e_1+e_2+e_3 \le \EEEmax$ is imposed on the 3\emph{N} matrix elements, where the $e_i$ are the principal quantum numbers of the single-particle harmonic-oscillator (HO) basis states.
To facilitate our calculations, we mainly use the normal-ordered two-body approximation (NO2B)~\cite{HaPa07,RoBi12} to the 3\emph{N} interaction, which was proven to be very accurate~\cite{HaPa07,RoBi12,BiPi13}.
%


\paragraph{Coupled-cluster method.}
For solving the many-body Schr\"odinger equation we employ a spherical formulation of CC theory~\cite{WlDe05,HaPa08,HaPa10,HaHj12}, which constitutes a good compromise between accuracy and computational efficiency.
In single-reference CC with singles and doubles excitations (CCSD)~\cite{PuBa82}, the ground state $|\Psi\rangle$ of a many-body Hamiltonian is parametrized by the exponential ansatz 
\mbox{$| \Psi \rangle = e^{T_1+T_2} \, | \Phi \rangle$},
where $T_n$ are $n$-particle-$n$-hole excitation operators 
acting on a single Slater-determinant reference state $| \Phi \rangle$.
Effects of the $T_3$ clusters are included through an \emph{a posteriori} correction to the energy via the CR-CC(2,3)~\cite{PiWl05,PiGo09,BiPi14} or the $\rm \Lambda CCSD(T)$~\cite{TaBa08a,Taba08b,BiPi13} method.
The underlying single-particle basis is a HO basis truncated in the principal oscillator quantum number \mbox{$2n+l \le \emax$}. We do Hartree-Fock (HF) calculations to optimize the single-particle basis, and perform the normal ordering with respect to the HF ground state.

%

\paragraph{Role of the three-body SRG model space.}
%
%

\begin{figure}[t]
\centering{ \includegraphics[width=1.0\columnwidth]{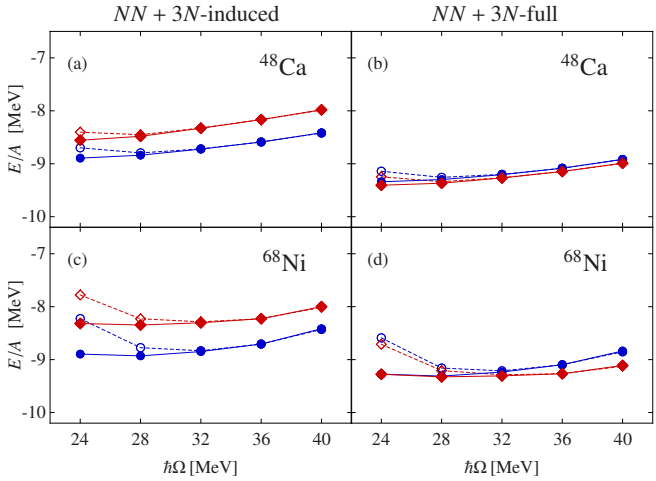}  \\[-10pt] 
}
\vspace{10pt}
\caption{
(Color online) Comparison of CCSD ground-state energies at flow parameters \mbox{$\alpha = 0.04\,\text{fm}^4$} (blue circles) and \mbox{$0.08\,\text{fm}^4$} (red diamonds), without (open symbols) and with (full symbols) frequency conversion,
using \mbox{$\EEEmax = 14$} and \mbox{$\emax = 12$}.
The frequency conversion was performed using the parent frequency \mbox{$\rm\hwSRG=36\,MeV$}.
}	
\label{fig:FrequencyConversion}
\end{figure}
%
%
The SRG evolution is performed in a finite model space and particularly for the evolution of the 3\emph{N} interaction, the model spaces  required to accurately represent the Hamiltonian become very large.
We parametrize our SRG model spaces by an angular-momentum dependent truncation $\ESRG(J)$ for the energy quantum numbers in the three-body Jacobi-HO basis in which the flow equation is solved~\cite{RoLa11,RoCa14}.
These parametrizations, referred to as \emph{ramps}, are defined by two plateaus of constant $\ESRG(J)$ with a linear slope  in between.
Earlier works employed ramp {\rampFourtyC} with 
\mbox{$\ESRG^{(\rampFourtyC)}(J\!\le\!\tfrac{5}{2})=40$} and 
\mbox{$\ESRG^{(\rampFourtyC)}(J\!\ge\!\tfrac{13}{2})=24$}~\cite{RoCa14,RoLa11,RoBi12,HeBi13,BiPi13,CiBa13,HeBo13}.
Already in medium-mass calculations, this ramp shows first deficiencies~\cite{BiLa13,HeBo13}. If the SRG evolution is performed at small frequencies $\hbar \Omega$, the momentum range covered in the truncated SRG model space is not sufficient to capture the relevant contributions of the initial Hamiltonian, resulting in an artificial increase of the ground-state energies.
We overcome this problem using the frequency conversion discussed in~\cite{RoCa14}, where we evolve the Hamiltonian at a sufficiently large frequency $\hwSRG$ and convert to the target frequency subsequently.
In Fig.~\ref{fig:FrequencyConversion} we show the $\hw$-dependence of CCSD ground-state energies obtained for ramp {\rampFourtyC} with and without frequency conversion.
This frequency conversion, used in all following calculations, eliminates the artificial increase of the energies at low frequencies and shifts the energy minima towards lower frequencies.
%
%
%
%
%
%
\begin{figure}[b]
\centering{ \includegraphics[width=1.0\columnwidth]{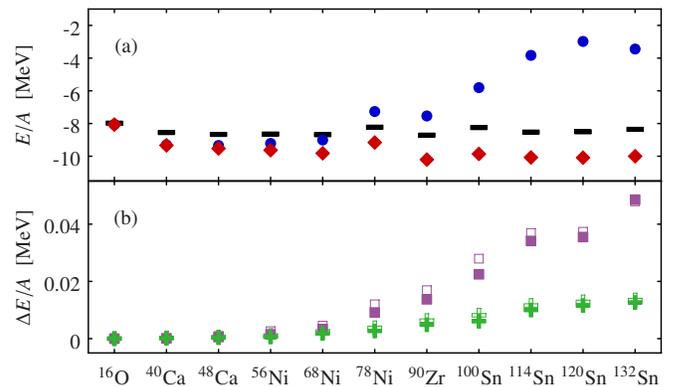}  \\[-10pt] 
}
\vspace{10pt}
\caption{
(Color online) 
(a) Comparison of CCSD ground-state energies corresponding to ramp {\rampFourtyC} (blue circles) and {\rampFourtyJ} (red diamonds) to experiment (black bars)~\cite{WaAu12}.
(b) Deviation of CCSD ground-state energies corresponding to ramp  {\rampFourtyK} (violet boxes) and {\rampFourtyL} (green crosses) from ramp {\rampFourtyJ} for the \NNN-induced (open symbols) and \NNN-full Hamiltonian (full symbols).
All calculations are performed 
for \mbox{$\EEEmax = 14$}, \mbox{$\alpha = 0.08 \, \rm fm^4$}, \mbox{$\rm\hw = 24\,MeV$} and \mbox{$\emax = 12$}.
}	
\label{fig:Ramps}
\end{figure}
%

Next we investigate the convergence with respect to the SRG model-space size.
To this end, we also employ a considerably larger model space defined by ramp {\rampFourtyJ}, with plateaus 
\mbox{$\ESRG^{(\rampFourtyJ)}(J\!\le\!\tfrac{7}{2})=40$} and 
\mbox{$\ESRG^{(\rampFourtyJ)}(J\!\ge\!\tfrac{11}{2})=36$}.
In Fig.~\ref{fig:Ramps}(a) we compare CCSD ground-state energies obtained for ramps {\rampFourtyC} and {\rampFourtyJ}.
For the lighter nuclei both ramps give very similar results, but
with increasing mass number we observe an increasing deviation.
For \elem{Ni}{56}, this deviation is about \mbox{0.4 MeV} per nucleon,
and grows to around 7 MeV per nucleon for the Sn isotopes. 
These results dramatically illustrate the importance of large SRG model spaces for heavier systems.
To assess the truncation errors related to ramp {\rampFourtyJ}
we introduce the two auxiliary 
ramps {\rampFourtyK} with  
\mbox{$\ESRG^{(\rampFourtyK)}(J\!\le\!\tfrac{7}{2})=40$} and 
\mbox{$\ESRG^{(\rampFourtyK)}(J\!\ge\!\tfrac{13}{2})=34$},
and {\rampFourtyL} with
\mbox{$\ESRG^{(\rampFourtyL)}(J\!\le\!\tfrac{5}{2})=40$} and 
\mbox{$\ESRG^{(\rampFourtyL)}(J\!\ge\!\tfrac{9}{2})=36$},
which probe the large-$J$ part of the 3\emph{N} SRG model space that is vital for heavier systems.
In Fig.~\ref{fig:Ramps}(b) we show the deviation of the CCSD ground-state energies for ramps {\rampFourtyK} and {\rampFourtyL} from the largest ramp {\rampFourtyJ}.
These deviations are below \mbox{50 keV} per nucleon even for the heaviest nuclei, which confirms convergence with respect to the SRG model-space size, and establishes ramp {\rampFourtyJ} as the standard used in the following.
We have also confirmed that the truncation in the low-$J$ part of the model space introduced only negligible errors.

%

\paragraph{CC convergence and triples correction.}
%
%

\begin{figure}[t]
\centering{ \includegraphics[width=1.17\columnwidth]{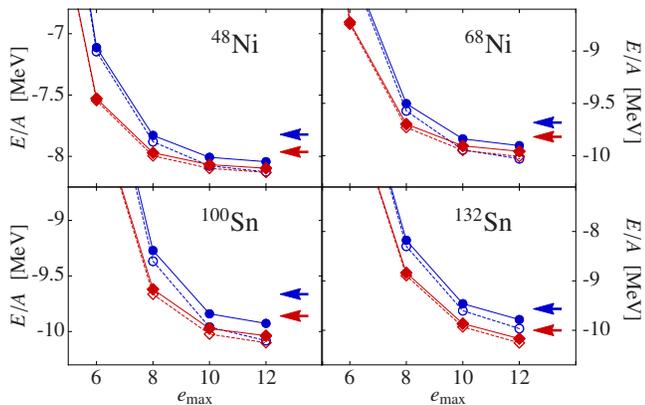}  \\[-10pt] 
}
\vspace{10pt}
\caption{
(Color online) 
Convergence of CR-CC(2,3) (full symbols) and $\Lambda$CCSD(T) (open symbols) ground-state energies for the \NNN-full Hamiltonian at  \mbox{$\alpha = 0.04\,\text{fm}^4$} (blue circles) and \mbox{$0.08\,\text{fm}^4$} (red diamonds), and with \mbox{$\EEEmax = 14$} and \mbox{$\hw$ = 24 MeV}.
Also shown are CCSD ground-state energies (arrows) from \mbox{$\emax = 12$} model spaces, where the upper (blue) arrows correspond to \mbox{$\alpha = 0.04\,\text{fm}^4$}. 
}	
\label{fig:eMax}
\end{figure}
%
%
Soft interactions allow for reasonably well converged CC calculations at \mbox{$\emax$ = 12}, as is apparent from Fig.~\ref{fig:eMax}, where we present ground-state energies from CCSD, $\Lambda$CCSD(T)~\cite{TaBa08a,Taba08b,HaPa10}, and CR-CC(2,3)~\cite{PiWl05,PiGo09,RoGo09,BiPi14}.
Both triples-correction methods are highly sophisticated and we note that the former can be obtained as an approximation to the latter~\cite{BiPi13}.
We observe noticeable differences for the \mbox{$\alpha$ = 0.04 $\rm fm^4$} interaction, where the magnitude of the triples correction itself is larger than for \mbox{$\alpha$ = 0.08 $\rm fm^4$}, with the $\Lambda$CCSD(T) results lying below their CR-CC(2,3) counterparts.
This is consistent with findings from quantum chemistry, where $\Lambda$CCSD(T) tends to overestimate the exact triples correction~\cite{Taube10}.
In the following, we use the size of the CR-CC(2,3) triples correction to estimate the rate of  convergence of the cluster expansion.
%

\begin{figure}[b]
\centering{ \includegraphics[width=1.0\columnwidth]{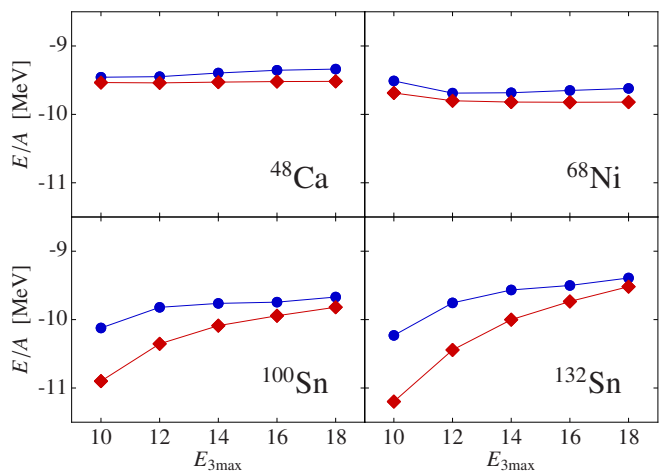}  \\[-10pt] 
}
\vspace{10pt}
\caption{
(Color online) 
Convergence of CCSD ground-state energies from $\emax$ = 12 CC model spaces, for the \NNN-full Hamiltonian at $\alpha=0.04\,\text{fm}^4$ (blue circles) and $0.08\,\text{fm}^4$ (red diamonds) with respect to $\EEEmax$.
Other parameters of the Hamiltonian as in Fig.~\ref{fig:eMax}.
}	
\label{fig:E3Max}
\end{figure}

\paragraph{Normal-ordering procedure.}
Because full matrix element sets with \mbox{$\EEEmax\approx16$} become inconveniently large~\cite{RoCa14}, 
we follow a procedure that avoids storage of full sets of \mbox{$\EEEmax>14$} matrix elements.
In a first step we perform a HF calculation including the complete 3\emph{N} interaction up to \mbox{$\EEEmax=14$} and use the HF ground state as reference for the normal-ordering of the 3\emph{N} interaction with the larger $\EEEmax$, where we selectively compute the subset of $JT$-coupled 3\emph{N} matrix elements~\cite{RoCa14} directly entering the normal-ordering.
Using the NO2B matrix elements we perform another HF calculation to obtain a reference state including the large-$\EEEmax$ information.
This process can be iterated until consistency is achieved, but a single iteration is typically sufficient.
In Fig.~\ref{fig:E3Max} we present CCSD ground-state energies of various nuclei using \mbox{$\EEEmax$ = 10} up to 18.
For the lighter nuclei \elem{Ca}{48} and \elem{Ni}{68}, convergence is reached around \mbox{$\EEEmax = 14$}. 
The situation changes for the heavier nuclei \elem{Sn}{100} and \elem{Sn}{132}, where the large values of $\EEEmax$ are in fact necessary to achieve  convergence.
%
%

%
The NO2B approximation is crucial since it allows to handle large values of $\EEEmax$.
However, earlier works show that for soft interactions contributions of the  residual normal-ordered 3\emph{N} interaction can become comparable to the triples correction~\cite{BiLa13,BiPi13}.
Most of these contributions stem from CCSD, while the residual 3\emph{N} interactions may be neglected in the triples correction~\cite{BiPi13}. 
Therefore, in the following we explicitly include the residual 3\emph{N} interaction up to  \mbox{$\EEEmax = 12$} when we solve the CCSD equations~\cite{HaPa07,BiLa13}, and use the NO2B matrix elements to cover the 3\emph{N} contributions up to \mbox{$\EEEmax = 18$}.
Particularly for the Ca and Ni isotopes, this practically eliminates the error of the NO2B approximation~\cite{BiLa13,BiPi13}.
An overall analysis of the sources of uncertainties present in our calculations suggests that for a given Hamiltonian at fixed $\alpha$, we obtain the energies with an accuracy of about 2\%.
%


\paragraph{Heavy nuclei from chiral Hamiltonians.}
%
%
%
%
\begin{figure*}[t]
\centering{ \includegraphics[width=\textwidth]{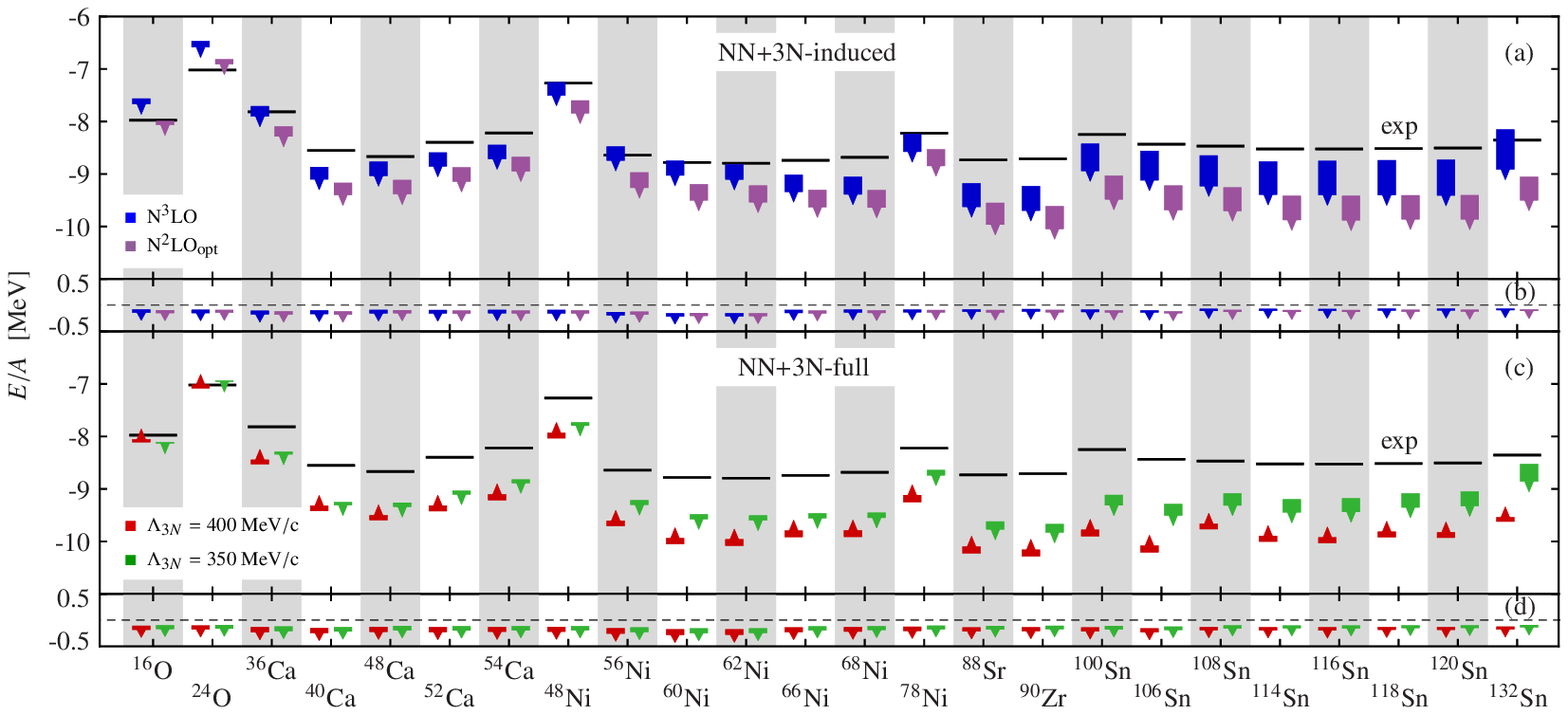}  \\[-10pt] 
}
\vspace{10pt}
\caption{
(Color online)
Ground-state energies from CR-CC(2,3) for 
(a) the \NNN-induced Hamiltonian starting from the N$^3$LO and N$^2$LO-optimized \emph{NN} interaction and 
(c) the \NNN-full Hamiltonian  with \mbox{$\Lambda_{3N}\rm = 400 \ MeV/c$} and  \mbox{$\Lambda_{3N} \rm= 350 \ MeV/c$}.
The boxes represent the spread of the results from \mbox{$\rm\alpha = 0.04 \ fm^4$ to $\rm\alpha = 0.08 \ fm^4$}, and the tip points into the direction of smaller values of $\alpha$.
Also shown are the contributions of the CR-CC(2,3) triples correction to the (b) \NNN-induced and (d) \NNN-full results.
All results employ \mbox{$\rm\hw = 24 \ MeV$} and 3\emph{N} interactions with \mbox{$\EEEmax = 18$} in NO2B approximation and full inclusion of the 3\emph{N} interaction in CCSD up to \mbox{$\EEEmax = 12$}.
Experimental binding energies~\cite{WaAu12} are shown as black bars.
}	
\label{fig:Overview}
\end{figure*}
%
%
%
%
The developments discussed above enable us to extend the range of accurate \emph{ab initio} calculations into the regime of heavy nuclei. In Fig.~\ref{fig:Overview} we present ground-state energies of closed sub-shell nuclei ranging from \elem{O}{16} to \elem{Sn}{132} for SRG-evolved chiral Hamiltonians with \mbox{$\EEEmax$ = 18} and for the two resolution scales \mbox{$\rm \alpha = 0.04 \, fm^4$} and \mbox{$\rm \alpha = 0.08 \, fm^4$} used to study the $\alpha$-dependence. In panels (a) and (c) we show ground-state energies obtained from CR-CC(2,3) in comparison to experiment, in panels (b) and (d) we depict the size of the triples correction beyond CCSD. 

First we consider the \NNN-induced results shown in Fig.~\ref{fig:Overview}(a). With increasing mass number, we observe a significant increase in the $\alpha$-dependence indicating growing contributions of SRG-induced 4\emph{N} (and multi-nucleon) interactions resulting from the initial \emph{NN} interaction. To confirm this trend, we show results starting from the optimized chiral \emph{NN} interaction $\rm N^2LO_{opt}$ presented in Ref.~\cite{EkBa13} in addition to the chiral $NN$ interaction at N$^3$LO of Ref.~\cite{EnMa03} used in all other calculations. Previous investigations have shown that when starting from a chiral \emph{NN} Hamiltonian, induced 4\emph{N} contributions are small for p- or lower sd-shell nuclei \cite{RoCa14,HeBi13,RoBi12}---this is confirmed within the truncation uncertainties by the present calculations. However, the effect of the omitted 4\emph{N} contributions is amplified when going to heavier nuclei and the $\alpha$-dependence indicates that these induced 4\emph{N} interactions are attractive. 

If we add the initial 3\emph{N} interaction to the chiral $NN$ interaction at N$^3$LO the picture changes. The $\alpha$-dependence of the \NNN-full Hamiltonian is significantly reduced compared to the \NNN-induced results, as seen in Fig.~\ref{fig:Overview}(c). In addition to the local 3\emph{N} interaction at N$^2$LO with initial cutoff \mbox{$\Lambda_{3N}$ = 400 MeV/c}, we employ a second cutoff \mbox{$\Lambda_{3N}$ = 350 MeV/c} for comparison \cite{RoCa14}. Our previous studies have shown that for both cutoffs, the induced 4\emph{N} interaction are small up into the sd-shell \cite{RoCa14,HeBi13}. For heavier nuclei, Fig.~\ref{fig:Overview}(c) reveals that the $\alpha$-dependence of the ground-state energies remains small for \mbox{$\Lambda_{3N}$ = 400 MeV/c} up to the heaviest nuclei. Thus, the attractive induced 4\emph{N} contributions that originate from the initial \emph{NN} interaction are canceled by additional repulsive 4\emph{N} contributions originating from the initial chiral 3\emph{N} interaction. By reducing the initial 3\emph{N} cutoff to \mbox{$\Lambda_{3N}$ = 350 MeV/c}, the repulsive 4\emph{N} component resulting for the initial 3\emph{N} interaction is weakened~\cite{RoCa14} and the attractive induced 4\emph{N} from the initial \emph{NN} prevails, leading to an increased $\alpha$-dependence indicating an attractive net 4\emph{N} contribution. All of these effects are larger than the truncation uncertainties of the calculations, such as the cluster truncation, as is evident by the comparatively small triples contributions shown in Fig.~\ref{fig:Overview}(b) and (d).

Taking advantage of the cancellation of induced 4\emph{N} terms for the \NNN-full Hamiltonian with \mbox{$\Lambda_{3N}$ = 400 MeV/c} we compare the energies to experiment. Throughout the different isotopic chains starting from Ca, the experimental pattern of the binding energies is reproduced up to a constant shift of the order of 1 MeV per nucleon. The stability and qualitative agreement of the these results over an unprecedented mass range is remarkable, given the fact that the Hamiltonian was determined in the few-body sector alone. 

When considering the quantitative deviations, one has to consider consistent chiral 3\emph{N} interaction at N$^3$LO, and the initial 4\emph{N} interaction.
In particular for heavier nuclei, the contribution of the leading-order 4\emph{N} interaction might be sizable. Another important future aspect is the study of other observables, such as charge radii. In the present calculations the charge radii of the HF reference states are systematically smaller than experiment and the discrepancy increases with mass. For \elem{O}{16}, \elem{Ca}{40}, \elem{Sr}{88}, and \elem{Sn}{120} the calculated charge radii are \mbox{$0.3$ fm}, \mbox{$0.5$ fm}, \mbox{$0.7$ fm}, and \mbox{$1.0$ fm} too small~\cite{WaAu12}. These deviations are larger than the expected effects of beyond-HF correlations and consistent SRG-evolutions of the radii. This  discrepancy will remain a challenge for future studies of medium-mass and heavy nuclei with chiral Hamiltonians.

%
%

\paragraph{Conclusions.}
In this Letter we have presented the first accurate \emph{ab initio} calculations for heavy nuclei using SRG-evolved chiral interactions. We have identified and eliminated a number of technical hurdles, e.g., regarding the SRG model space, that have inhibited state-of-the-art medium-mass approaches to address heavy nuclei. As a result, many-body calculations up to \elem{Sn}{132} are now possible with controlled uncertainties on the order of 2\%. 
The qualitative agreement of ground-state energies for nuclei ranging from \elem{O}{16} to \elem{Sn}{132} obtained in a single theoretical framework demonstrates the potential of ab initio approaches based on chiral Hamiltonians. 
This is a first direct validation of chiral Hamiltonians in the regime of heavy nuclei using \emph{ab initio} techniques.
Future studies will have to 
involve consistent chiral Hamiltonians at N$^3$LO considering initial and SRG-induced 4\emph{N} interactions and provide an exploration of other observables.

\paragraph{Acknowledgements.}
We thank Piotr Piecuch for helpful discussions and Petr Navr\'atil for providing us with the \textsc{ManyEff} code~\cite{NaKa00}.
Supported by the
Deutsche Forschungsgemeinschaft through contract SFB 634,
by the Helmholtz International Center for FAIR (HIC for
FAIR) within the LOEWE program of the State of Hesse, and
the BMBF through contract 06DA7047I. Numerical calculations
have been performed at the computing center of the TU
Darmstadt (lichtenberg), at the J\"ulich Supercomputing Centre
(juropa), at the LOEWE-CSC Frankfurt, and at the National
Energy Research Scientific Computing Center supported by
the Office of Science of the U.S. Department of Energy under
Contract No. DE-AC02-05CH11231.

		
\end{document}